\documentclass[twocolumn,amsmath,amssymb,superscriptaddress,pra]{revtex4-1}

\usepackage{color}
\usepackage{graphicx}
\usepackage{amssymb}
\usepackage{amsmath}
\usepackage{mathrsfs}
\newcommand{\ud}{\mathrm{d}}

\graphicspath{ {./figures/} {./}}
\newcommand\figref{Fig.~\ref}

\begin{document}

\title{Stable integrated hyper-parametric oscillator based on coupled optical microcavities}

\author{Andrea Armaroli}
\email{andrea.armaroli@enssat.fr}
\affiliation{FOTON (CNRS-UMR 6082), Universit\'{e} de Rennes 1, ENSSAT, 6 rue de Kerampont, CS 80518, 22305 Lannion CEDEX, France}
\author{Patrice Féron}
\affiliation{FOTON (CNRS-UMR 6082), Universit\'{e} de Rennes 1, ENSSAT, 6 rue de Kerampont, CS 80518, 22305 Lannion CEDEX, France}
\author{Yannick Dumeige}
\affiliation{FOTON (CNRS-UMR 6082), Universit\'{e} de Rennes 1, ENSSAT, 6 rue de Kerampont, CS 80518, 22305 Lannion CEDEX, France}

\begin{abstract}
We propose a flexible scheme based on three coupled optical microcavities which permits to achieve stable oscillations in the microwave range, the frequency of which depends only on the cavity coupling rates. We find that the different dynamical regimes (soft and hard excitation)  affect the oscillation intensity but not their period. This configuration may permit to implement compact hyper-parametric  sources on an integrated optical circuit, with interesting applications in communications, sensing and metrology.
\end{abstract}

\maketitle

Nonlinear effects in optical microresonators provide fundamental functionalities for high speed all-optical signal processing \cite{Ilchenko2006}. Frequency conversion \cite{Turner2008}, switching \cite{Ibanescu2002}, signal regeneration in communications \cite{Ghisa2007} and  optical generation of microwaves \cite{Matsko2005b,Duchesne2010} are examples of applications which have attracted a fair deal of attention in the latest years.

The generation of oscillations at microwave frequency ranging from 10 GHz to 200 GHz based on the beating of optical oscillations is an enabling technology not only for high-speed communications (e.g.~in aerospace industry), but also for metrology, optical clocks and sensing. One of the most interesting proposals is to exploit a set of adjacent resonances of an optical microresonator  \cite{Matsko2005b}. The ubiquitous four-wave mixing (FWM) between nearly equi-separated resonant modes manifests itself, at high enough power, in the conversion from a quasi continuous excitation to a frequency comb \cite{Del'Haye2009,Chembo2010,Cazier2013,Soltani2012}, and ultimately forms train of pulses and solitary waves inside the cavity \cite{Herr2014a}. The threshold power to observe such phenomena depends on the nonlinearity of the medium and on the cavity lifetime, or quality factor $Q$, which should be above $Q\approx10^7$. This poses strong technological constraints and octave-spanning frequency combs are generally observed in  large diameter ($\approx 200 \mu \mathrm{\!m}$) glass or crystalline microresonators. Thus, due to the low nonlinearity, the required power levels causes generally thermal dissipation concerns. Finally as the frequency spacing of resonant modes, i.e.~the free spectral range (FSR), corresponds to the microwave oscillation frequency, this implies that, in order to obtain oscillations at e.g.~$10$GHz, a microresonator of more than $1\,$mm radius is needed. This clearly poses a serious limit to integration.
It would be beneficial to scale this technology to an integrated optical platform, based on semiconductors (Si or III-V alloys), which exhibits much stronger nonlinearities. There are two limits: usually it is much harder to obtain $Q>10^6$, due to the technological processes involved in fabrication and, because of the small size, the FSR is in the THz range.

In order to obtain oscillations in the GHz range in a single optical microcavity (microring, microsphere, or photonic crystal cavity), the usual approach is to couple to an additional degree of freedom, such as a time-delayed nonlinear response \cite{Malaguti2011c,Armaroli2011a,Vaerenbergh2012}. Alternatively  a system of coupled cavities can be designed. This last solution is normally limited to two cavities and the system is destabilized and starts to oscillate at a frequency which is a sort of beating between the resonances of the overall system \cite{Maes2009,Grigoriev2011a,Dumeige2015}. The main disadvantage is that the pulsation period is of the same order of the cavity lifetime, so that to obtain oscillations in the GHz range (i.e.~a period of $0.1\,$ns) we are limited to a $Q$-factor of about $10^4$ which in turn implies huge power levels. Finally in both approaches (delayed response or a pair of coupled cavities, or even both), the pulsations (limit cycles) are subject to a period doubling cascade to chaos. 

In our work we propose a system of three evanescently-coupled optical micro-race-track resonators with instantaneous Kerr response. By supposing to be able to tune the cavity coupling rate independently from the resonance frequency and lifetime, we obtain that in the limit of large coupling the system exhibits stable oscillations in the GHz range, the frequency of which is largely independent of power and not subject to a chaotic evolution. Thus we can implement this configuration in a semiconductor platform, with large  $Q\approx10^5$ values and accessible power levels. This promises to be a flexible integration strategy for obtaining a microwave oscillator on an optical integrated cicuit.
We consider a system composed by evanescently coupled optical microcavities (single mode or with a large FSR), the time evolution of which reads, in dimensional units, as \cite{HausBook,Fan2003,Grigoriev2011a,Abdollahi2014,Dumeige2015}
\begin{equation}
\begin{split}
\label{eq:dim}
\frac{\ud{A}_j}{\ud T}=\left[i(\tilde\delta_j+\tilde\chi_j\left|A_j\right|^2)-\frac{1}{\tilde \tau_j}\right]A_j \\+i \sum_{k\neq j}{\tilde\gamma_{jk} A_k}+ \sqrt{\frac{2}{\tilde\tau_{wg}}} s_{in}\delta_{j1}
\end{split}
\end{equation}
where $j,k=1-3$, $\tilde\delta_j=\tilde\omega_j-\tilde\omega_L$ is the detuning of the laser excitation from the $j$-th cavity resonance frequency, $\tilde \tau_j$ is the cavity lifetime, $\tilde\tau_{wg}$ quantifies the coupling from the input waveguide to the first cavity. { We further assume, for the sake of simplicity, that the decay into the waveguide is negligible with respect to the intrinsic cavity contribution, i.e.~$\tilde \tau_1\ll\tilde\tau_{wg}$  (undercoupling), as opposed to the  critically coupled (the escape and decay rates are equal) or overcoupled (the escape in the waveguide is dominant) case. }
$A_j$ are normalized such that the square amplitude is the energy stored in the cavity (in J), $|s_{in}|^2$ is the power in W in the external waveguide coupled into the first cavity, $\tilde\gamma_{jk}=\tilde\gamma_{kj}$ are the coupling rate of cavity $j$ and $k$ and are assumed to be real. The independent variable $T$ represents the time expressed in seconds.
We pose the effective nonlinear coefficient $\tilde\chi_j=\frac{\tilde\omega_j c n_2}{n_\mathrm{eff}^2\mathcal{V}}$, where $n_2$ is the Kerr coefficient, $n_\mathrm{eff}$ is the modal effective index and $\mathcal{V}$ is the modal effective volume. 

\begin{figure}
	\centering
				\includegraphics[width=0.40\textwidth]{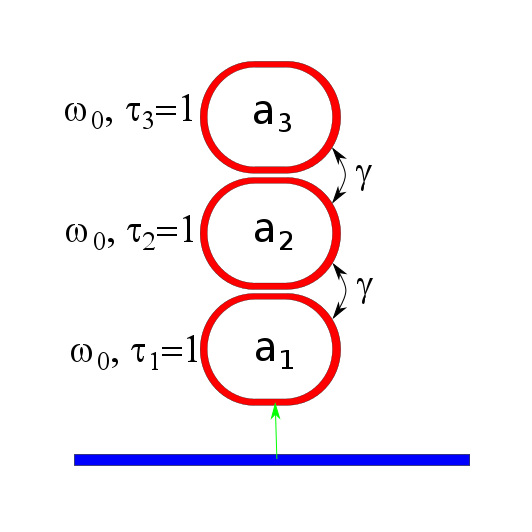}	
	\caption{Configuration of the oscillator based on three coupled micro-resonators, evanescently excited by an external under-coupled waveguide.}
	\label{fig:scheme}
\end{figure}

In the  three cavity configuration of \figref{fig:scheme} the coupling coefficients, lifetimes and modal properties are assumed to be the same for each cavity $\tilde\tau_j=\tilde\tau$, $\tilde\omega_j=\tilde\omega$, $\tilde\delta_j=\tilde\delta$, $\tilde\chi_j=\tilde\chi>0$ for $j=1,2,3$, $\tilde\gamma_{jk}=\tilde\gamma$ for $|j-k|=1$ and $\tilde\gamma_{jk}=0$ otherwise. By introducing the normalization $a=A/\sqrt{I_0}$, $t=T/\tilde\tau$, with $I_0=(\tilde\tau \tilde\chi)^{-1}$, we derive from \eqref{eq:dim} the following adimensional model
\begin{equation}
\begin{aligned}
\frac{\ud{a}_1}{\ud t}&=\left[i(\delta+\chi\left|a_1\right|^2)-1\right]a_1 +i \gamma a_2+ \sqrt{P}\\
\frac{\ud{a}_2}{\ud t}&=\left[i(\delta+\chi\left|a_2\right|^2)-1\right]a_2 +i \gamma a_1+ i \gamma a_3\\
\frac{\ud{a}_3}{\ud t}&=\left[i(\delta+\chi\left|a_3\right|^2)-1\right]a_3 +i \gamma a_2
\end{aligned}
\label{eq:3adim}
\end{equation} 
moreover $\gamma=\tilde\gamma\tilde\tau_j$, $\delta=\tilde\delta\tilde\tau$ and $P=2\tilde\tau^2 |s_{in}|^2/(I_0\tilde\tau_{wg})=2\tilde\tau^3 |\tilde\chi||s_{in}|^2/\tilde\tau_{wg}$ is the actual power coupled in the first cavity.  
\textcolor{black}{The equilibria (steady states) of \eqref{eq:3adim} are found by imposing $\ud a_j/\ud t=0$, for $j=1,2,3$.}

In the linear limit ($\chi=0$) and free evolution ($P=0$), we can easily obtain the linear modes of the structures, \textcolor{black}{by looking for the eigenvalues $\delta$ \textcolor{black}{(the real and imaginary parts of which correspond to frequency and lifetime, respectively)} and  eigenmodes $u = [a_1,a_2,a_3]^T$ at equilibrium}; the cavity modes split into a central resonance $\delta_0 = 0-i$ with eigenmode $u_0 = (\frac{1}{\sqrt{2}},0,-\frac{1}{\sqrt{2}})^T$ and a pair of symmetric sidebands $\delta_{\pm 1}=\pm\sqrt{2}|\gamma|-i$ with eigenmodes $u_{\pm 1} = \left(\frac{1}{2},\pm\frac{\sqrt{2}}{2},\frac{1}{2}\right)^T$, \textcolor{black}{ i.e.~the lifetime is the same for the three modes and $a_2$ is null (resp.~dominant) for the central (resp.~lateral) mode.}


In the nonlinear case ($\chi=1$), if $\gamma\approx1$ is considered, like in \cite{Maes2009,Grigoriev2011a,Dumeige2015}, we obtain a complicated bifurcation diagram and different regimes of self-pulsing, which may be subject to period-doubling bifurcation. This is due to fact that the resonances of the coupled system are too close, they are thus all significantly excited by the external source and this causes an intrinsic instability of the beating among them.

In the limit of  $\gamma\gg 1$, the three resonances are widely split and each mode can be selectively excited: if the central mode is chosen, the Kerr effect leads to the conversion of photons into the lateral modes and  a stable limit cycle of period $T=2\pi/\delta_{\pm 1}=\sqrt{2}\pi/\gamma$ is expected to appear.
In this limit we can approximate the steady state of \eqref{eq:3adim} as $a_1^E=-a_3^E$ and $a_2^E=0$, which corresponds to the linear eigenmode $u_0$ and is justified also by the strong coupling, which inhibits the energy storage in the middle cavity and we can express implicitly the solution for the intensity $I_1=|a_1^E|^2=I_3=|a_3^E|^2$ as 
\begin{equation}
	4 I_3(1+(\delta+\chi I_3)^2)= P
\label{eq:staticcurve}
\end{equation}
which has the same form of the bistability curve for the single nonlinear optical cavity, except for the factor 4, which reflects the partition of energy between cavities 1 and 3.
As in the conventional bistable cavity, the system is thus monostable if $\delta\ge-\sqrt{3}$ and exhibits multiple solutions if $\delta<-\sqrt{3}$. 
It is well-known that a pair of saddle-node (SN) bifurcations (annihilation of one stable and one unstable equilibrium) appear for 
\begin{equation}
			I_3^{\pm}=\frac{-2 \delta \pm\sqrt{-3+\delta ^2}}{3 \chi}.
			\label{eq:LP}
\end{equation}
The interaction of modes manifests itself as Hopf (H) bifurcations, where a limit cycle coexist locally with an equilibrium. 
An elementary but tedious analysis permits to estimate the eigenvalues of the Jacobian matrix associated to \eqref{eq:3adim} and locate the two H bifurcations (where a pair of conjugate imaginary  eigenvalues appears):
\begin{equation}
	I_3^{H\pm} = - \frac{4\delta}{3\chi} \pm \frac{\sqrt{4\gamma^2(\delta^2-3)-6}}{3\gamma\chi}.
	\label{eq:Hopfpoints}
\end{equation}
We notice that those points exist only in the multistable case and we can have two different scenarios: (i) $I_1^{H-}>I_1^+$, if $-3\sqrt{\frac{3}{5}}<\delta<-\sqrt{3}$, and (ii) $I_1^-<I_1^{H-}<I_1^+$, if  $\delta<-3\sqrt{\frac{3}{5}}$.
This estimates provide important indications on the most interesting regions in the space of parameters. 

\begin{figure}
	\centering
		\includegraphics[width=0.32\textwidth]{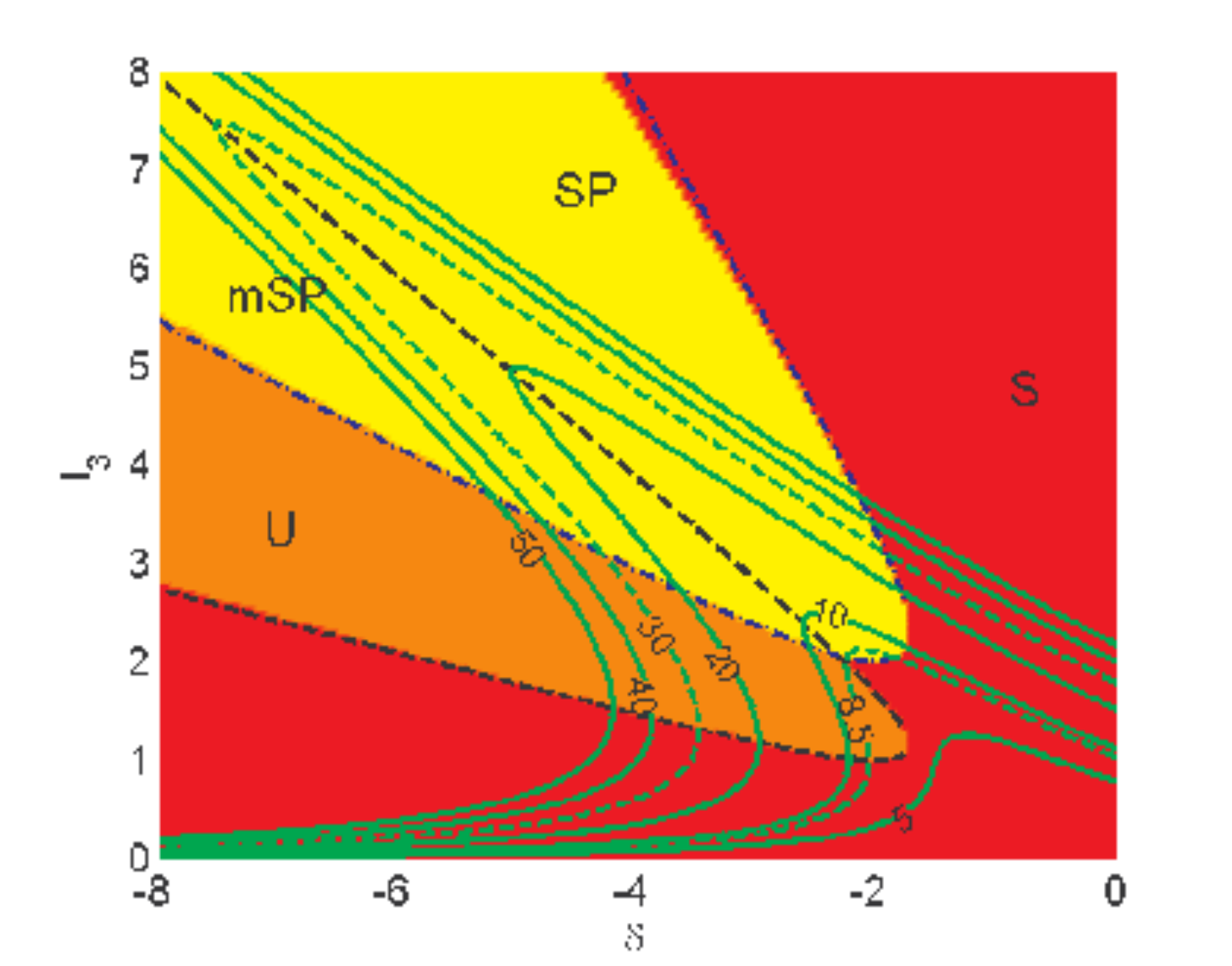}	
	\caption{Stability and self pulsation in the $(\delta,I_3)$ plane; the different shaded regions represent stable (S), unstable (U), self-pulsing (SP) and (mSP) multistable SP numerically computed equilibria; the black dashed line correspond to the analytical estimate SN bifurcations of \eqref{eq:LP}, the blue dash-dotted lines to the H bifurcations, \eqref{eq:Hopfpoints}. The mSP region is located between the lower blue dashed-dotted and the upper black dashed line; the green level curves represent the power level corresponding to each point in the plane, i.e.~the nonlinear frequency response obtained by solving \eqref{eq:staticcurve} for a fixed value of $P$; the dashed green curves show the values which we study in detail in \figref{fig:bifurcPlow} and \figref{fig:bifurcPhigh}.}
	\label{fig:parammap}
\end{figure}

To confirm the validity of our approximated relations, we now  set $\gamma=40$ and resort to detailed numerical calculations; we set the power $P$ and use $\delta$ as the bifurcation parameter.
First we classify the different dynamical regimes in the $(\delta,I_3)$ plane. It proves  straightforward, see \cite{Dumeige2015}, to fix $I_3$ and solve numerically the nonlinear system for equilibria  backwards to obtain $I_{1,2}$ and finally $P$; for each point we compute numerically the eigenvalues of the Jacobian. 
The results are reported in \figref{fig:parammap}. For low power $P<8$, analogously to a single Kerr cavity, the system is mono- or bistable. Then, in a small interval $8<P\lesssim 8.81$, a pair of H bifurcations exist, which occur in the upper bistable branch: they are both supercritical and we define this as the slow regime. Finally, for larger power level, the first H bifurcation is found between the two SN points, while the second is still on the upper branch. The low energy one is subcritical, while the high-energy one is still supercritical. The limit cycles are now stronger attractors than in the slow regime above and we define it as the fast regime. We include in \figref{fig:parammap} also our analytical estimates which are  in excellent agreement to the numerically obtained values.


\begin{figure}[h]
	\centering
		\includegraphics[width=0.32\textwidth]{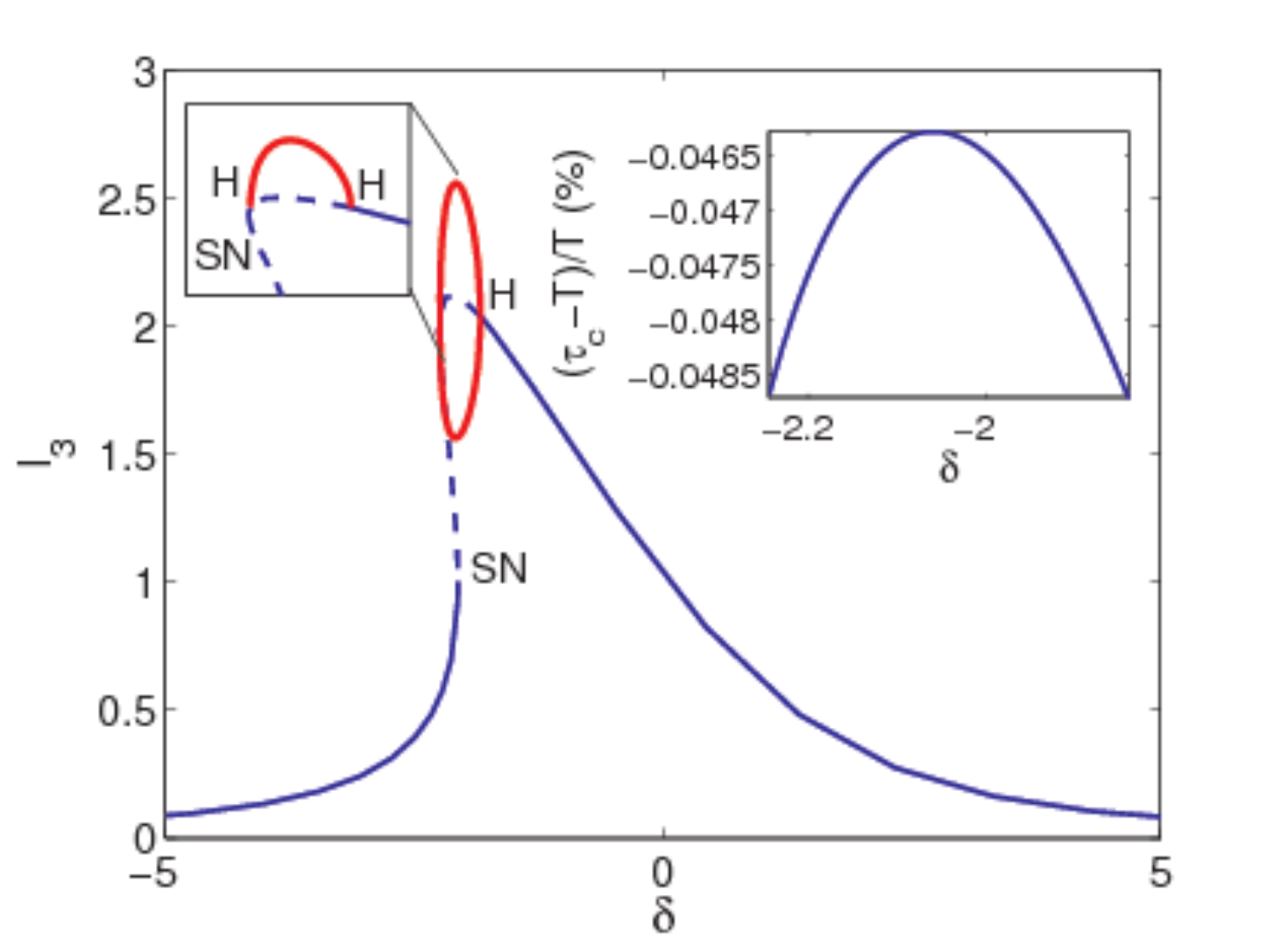}
	\caption{Bifurcation diagram of the intensity $I_3$ varying the laser detuning $\delta$, for $\gamma=40$ and $P=8.5$. The blue line represents the bifurcation of equilibria and the red line represents the bifurcation of limit cycles (maxima and minima of each cycle); a solid line is a stable solution, while a dashed line an unstable one: bifurcation points are labeled as SN (saddle-node) and H (Hopf). The solution of \eqref{eq:staticcurve} and its intersections with \eqref{eq:LP} and \eqref{eq:Hopfpoints} correspond perfectly to the numerical solutions.The top right inset shows the relative deviation of the period of limit cycles from $T$ (the beating period between lateral sidebands) as a function of $\delta$, the left one a zoom on the details of the bifrucation diagram (with only the limit cycle maxima). }
	\label{fig:bifurcPlow}
\end{figure}

\begin{figure}
	\centering
		\includegraphics[width=0.32\textwidth]{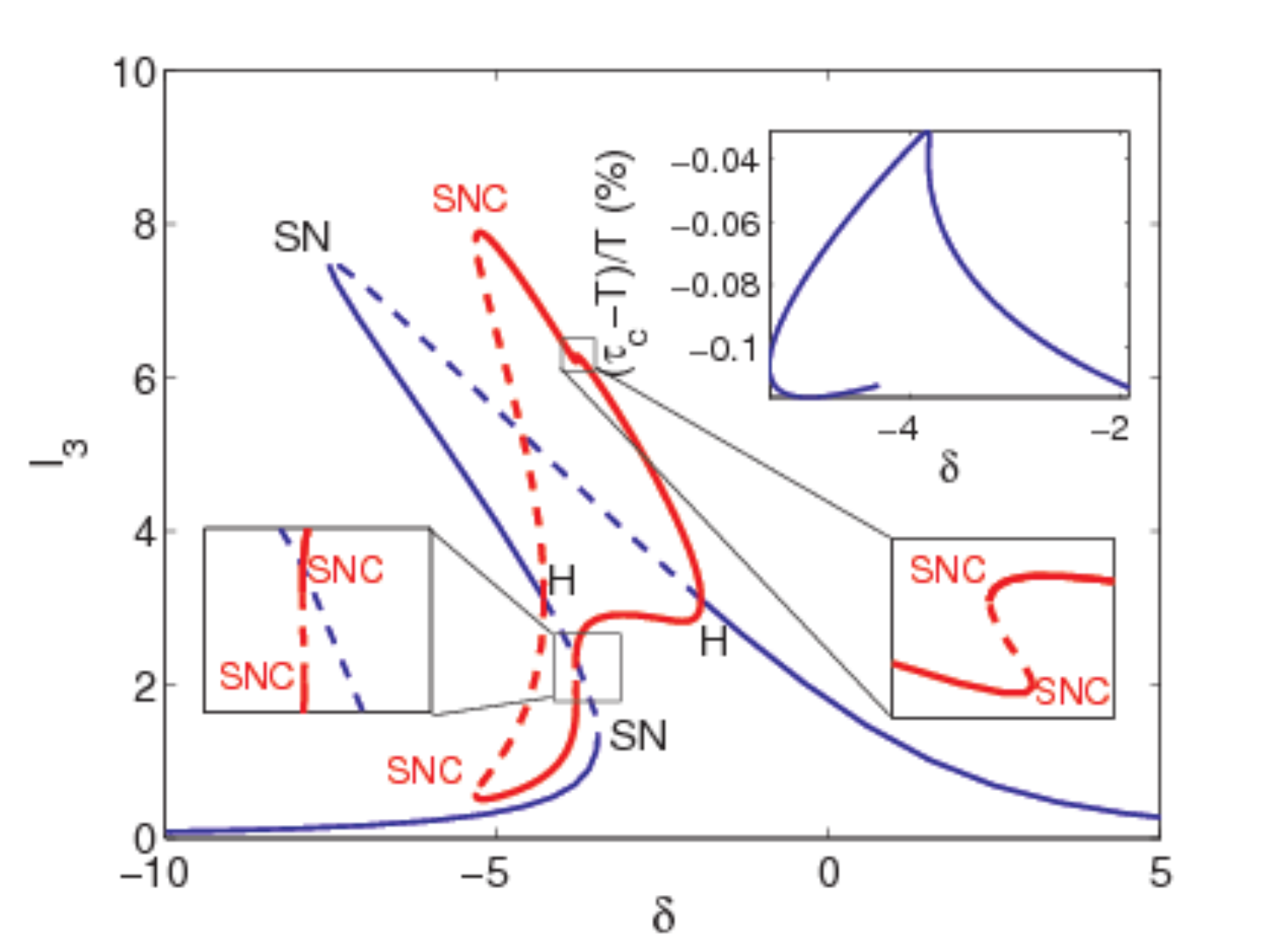}
	\caption{Same as Fig.~\ref{fig:bifurcPlow} for $P=30$. Notice that the limit cycles  also exhibit multistability, stable and unstable branches annihilate in SNC (saddle-node of limit cycles) bifurcations. A branch of stable limit cycles coexist with a low intensity stable equilibrium, and requires hard excitation, a second branch of stable limit cycles appears at larger detuning and exhibits soft excitability.}
	\label{fig:bifurcPhigh}
\end{figure}

\begin{figure}
	\centering
		\includegraphics[width=0.35\textwidth]{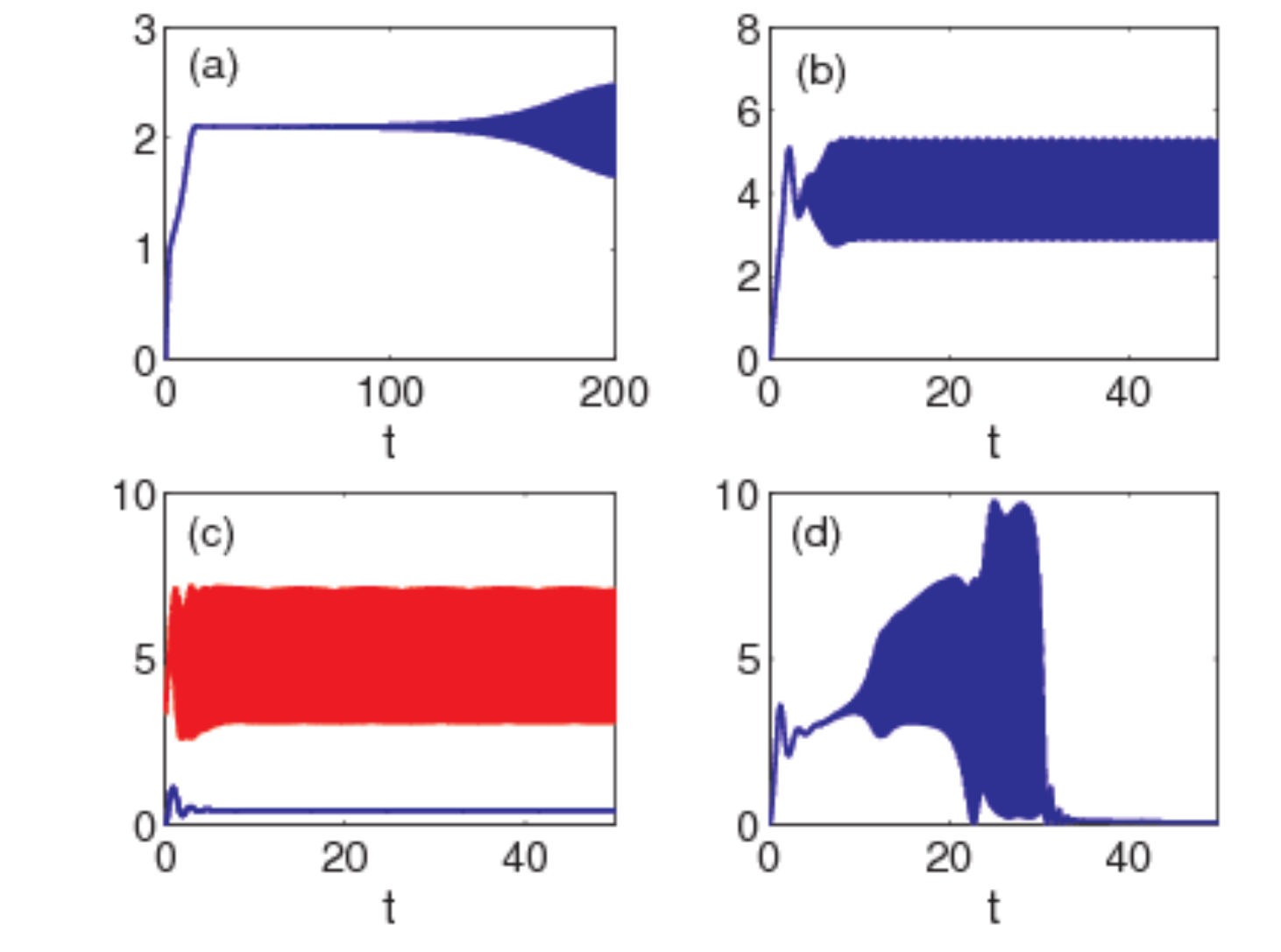}
	\caption{Time evolution of $I_3$ from the numerical solutions of \eqref{eq:3adim}, for  $\gamma=40$. (a)  $\delta=-2$, $P=8.5$; (b) $\delta=-3$, $P=30$; (c) $\delta=-4.5$, $P=30$: the blue line corresponds to soft excitation (noise initial conditions), the red one to hard excitation ($a_1(0)=-a_3(0)=2$);  (d) $P=30$, $\delta$ varies from -1 to -11.}
	\label{fig:timeevolution}
\end{figure}

Then we study in detail the bifurcation of equilibria and limit cycles as a function of $\delta$ for $P=8.5$ (Fig.~\ref{fig:bifurcPlow}) and $P=30$ (Fig.~\ref{fig:bifurcPhigh}) by means of the Matcont toolbox \cite{Dhooge2003}.
In the first case, Fig.~\ref{fig:bifurcPlow}, the two saddle node bifurcation occur before the two H bifurcations. These latter are both supercritical and a branch of stable limit cycles connects them. At this low power level, the equilibrium is only slightly unstable and self-pulsation requires many cavity lifetimes to be achieved, see Fig.~\ref{fig:timeevolution}(a).
The second situation is much richer, \figref{fig:bifurcPhigh}: the first H bifurcation occurs in the unstable branch of the bistable curve, it is subcritical (i.e.~it gives rise to an unstable limit cycle). Then the (now stable) equilibrium undergoes the second SN bifurcation and recovers its stability at the second (supercritical) H point. The limit cycles are still on a curve connecting the two H bifurcations, but in this case they exhibit themselves bistability (or saddle-node bifurcations of limit cycles, SNC). It is interesting to notice that the first (mostly detuned) stable branch of pulsating solutions has large amplitudes and coexists closely to the low energy stable equilibrium. The second stable branch exhibits decreasing oscillation amplitude which vanishes at the second H point. There is a crucial difference between the two branches: while the second one is achieved starting from a noise in the cavity system (soft excitation, see Fig.~\ref{fig:timeevolution}(b)) the first branch demands a finite energy inside the cavity (hard excitation, see Fig.~\ref{fig:timeevolution}(c))  and thus can be observed only by progressively decreasing the detuning, as plotted in Fig.~\ref{fig:timeevolution}(d) where the different dynamical regimes are shown to be accessible by sweeping $\delta$. 

At each point, we obtain also the period of oscillation (its relative deviation from $T$ as a function of $\delta$ is shown in the insets of Figs.~\ref{fig:bifurcPlow}--\ref{fig:bifurcPhigh}): it comes up that in any of the above cases, the period is virtually locked at $T=\sqrt{2}\pi/\gamma$, which correspond exactly to the oscillations between the linear resonances obtained above; the deviations from this value are of the order of $1/\gamma^2$. Thus we proved the existence of different  self-pulsation regimes, the frequency of which is the same irrespective of the power and detuning level. They are also stable, in the sense that in the whole SP region of Fig.~\ref{fig:parammap} the cycles never undergo a period-doubling cascade, as it occurs in \cite{Maes2009,Armaroli2011a}. 

We finally comment on the physical accessibility of this approach. We assume to operate at $\lambda=1.55\,\mu\mathrm{\!m}$ and take $\tilde\tau_j=\tilde\tau=1\,$ns, so that the cavity has a \textcolor{black}{$Q=\frac{\tilde\omega\tilde\tau}{2}\approx 6.1 \times 10^5$}. Consider a racetrack cavity with minimum curvature radius $R=10\,\mu\mathrm{\!m}$ and mode area $A_\mathrm{eff}=1\,\mu{\rm\!m^2}$ (upper bound): the modal volume is $\mathcal{V} \approx 0.63\,\mu\mathrm{m^3}$.
and the effective index of the mode is $n_\mathrm{eff}=2$. Notice that the resulting round-trip time in the cavity is still much shorter than the lifetime, thus Ikeda instabilities would require power and detuning values incompatible with the present analysis.
The medium is a semiconductor of refractive index $3.48$, Kerr index $n_2=2\times 10^{-17}\,\mathrm{m^2/W}$ \cite{Wagner2009}, thus we get $\tilde\chi=2.90\times 10^{22}\,\mathrm{[J s]^{-1}}$.
The scaling intensity results $I_0 =34.4\,$fJ.
We also assume weak waveguide coupling $\tilde\tau_{wg}=10\tilde\tau$, so that the first cavity is undercoupled to the waveguide and the quality factor does not vary considerably from a cavity to the other.  
With this values, $P=8.5$ corresponds to a power in the waveguide $|s_{in}|^2=1.5\,$mW and $P=30$ to $|s_{in}|^2=5.2\,$mW. \textcolor{black}{These power levels are feasible despite the undercoupling regime.}
As far as the coupling is concerned, a basic modal calculation, \cite{Little1997,HausBook}, permits to estimate that two waveguides of cross section $400\times300\,$nm with a gap of $200\,$nm require only about a coupling length $L_{cpl}=4\,\mu\mathrm{\!m}$ to achieve the normalized value of $\gamma=40$.
This value was chosen in order to obtain an oscillation frequency  of about 9 GHz. Moreover we numerically investigate if the exact match of the resonant frequencies and coupling coefficients is crucial to observe these oscillations. We verified numerically that quite a strong mismatch (in the order of $1\%$, in each parameter) reflects in the case of  Fig.~\ref{fig:bifurcPhigh} in less than $0.5\%$ change in the period, not affecting significantly the structure of the bifurcation diagram. This proves that our approach permits to generate microwave hyper-parametric oscillations and can be implemented on a standard integrated optical platform. \textcolor{black}{We remark that to this aim the undercoupling regime is less demanding than a choice relying upon critical waveguide-cavity coupling, because only the two gaps between resonators are critical.}

In summary, we proposed a design, based on a triad of evanescently coupled racetrack microcavities, which permits to achieve hyper-parametric oscillations at a frequency depending only on the coupling rate among the cavities. The oscillations are stable and exist in a wide region of the parameter space. We found different excitation regimes (slow and fast, hard and soft) and prove the robustness of the design. This system can be easily implemented in an up-to-date semiconductor platform for optical integrated circuits  \cite{Barwicz2006,Duchesne2010,Pu2015}, but also in more advanced settings such as photonic crystal nanocavities,which are virtually monomodal and can exhibit very high $Q$-factors. This may permit to achieve compact microwave sources on an optical chip, operating in a frequency range inaccessible before, with important applications in telecommunications, sensing and metrology. 

\textbf{Funding.} Y.~D. acknowledges the support of the Institut Universitaire de France (IUF).
\end{document}